\documentclass[prd,twocolumn,aps,amsmath,amssymb,nofootinbib,preprintnumbers]
{revtex4}
\IfFileExists{srcltx.sty}{\usepackage[active]{srcltx}}%

\usepackage{graphicx}
\usepackage{dcolumn}
\usepackage{bm}
\usepackage{amsmath}
\usepackage{amsfonts}

\def\ls{\mathrel{\lower4pt\vbox{\lineskip=0pt\baselineskip=0pt
           \hbox{$<$}\hbox{$\sim$}}}}
\def\gs{\mathrel{\lower4pt\vbox{\lineskip=0pt\baselineskip=0pt
           \hbox{$>$}\hbox{$\sim$}}}}
\def\drawbox#1#2{\hrule height#2pt

\hbox{\vrule width#2pt height#1pt \kern#1pt
              \vrule width#2pt}
              \hrule height#2pt}

\def\Asym#1#2{\vcenter{\vbox{\drawbox{#1}{#2}
              \kern-#2pt       
              \drawbox{#1}{#2}}}}

\newcommand{\beq}{\begin{equation}}
\newcommand{\eeq}{\end{equation}}

\begin{document}

\title{Sub-eV Hubble scale inflation within gauge mediated
supersymmetry breaking}

\author{Rouzbeh Allahverdi$^{1}$}
\author{Asko Jokinen$^{2, 3}$}
\author{Anupam Mazumdar$^{3}$}

\affiliation{$^{1}$~Perimeter Institute for Theoretical Physics, Waterloo, 
ON, N2L 2Y5, Canada \\
$^{2}$~Laboratoire de Physique Th\'eorique et Astroparticules, CNRS UMR 5207, 
Universit\'e Montpellier II, F-34095 Montpellier Cedex 5, France.\\
$^{3}$~NORDITA, Blegdamsvej-17, Copenhagen-2100, Denmark}

\begin{abstract}
Minimal Supersymmetric Standard Model with gauge mediated
supersymmetry breaking has all the necessary ingredients for a
successful sub-eV Hubble scale inflation $H_{\rm inf} \sim
10^{-3}-10^{-1}$ eV. The model generates the right amplitude for
scalar density perturbations and a spectral tilt within the range,
$0.90 \leq n_s \leq 1$.  The reheat temperature is $T_{\rm R} \ls 10$
TeV, which strongly prefers electroweak baryogenesis and creates the
right abundance of gravitinos with a mass $m_{3/2} \gs 100$ keV to be
the dark matter.
\end{abstract}

\maketitle

\noindent

Embedding a very low scale inflation within a particle physics model
is a challenging problem.  It is not only difficult to obtain
sufficient number of e-foldings, right amplitude for the scalar
density perturbations, the right tilt in the power spectrum, but also
generating baryon asymmetry and dark matter simultaneously.

Recently there has been a real progress in our understanding of
embedding inflation within particle physics, particularly within the
Minimal Supersymmetric Standard Model (MSSM), where the inflaton
belongs to the MSSM instead of being an {\it ad-hoc} gauge
singlet. Foremost the models~\cite{AEGM,AKM} not only predict the
right amplitude of the scalar density perturbations and the tilted
spectrum, see also~\cite{AM}, but are also testable at
LHC~\cite{AEGJM}~\footnote{The inflaton candidates are,
$LLe,~udd$~\cite{AEGM} and $NH_uL$~\cite{AKM} flat directions. Here
$L$ denotes left-handed sleptons, $u,~d,~e$ denote right-handed
squarks and sleptons, respectively, $N$ denotes right-handed
sneutrinos and $H_u$ is the Higgs which gives masses to the up type
quarks. The $LLe,~udd$ mass should be $\geq 340$~GeV~\cite{AEGM},
which is within the reach of LHC.  For other attempts of inflation
with flat directions, see~\cite{FEW,LYTH1}. For a review on MSSM flat
directions, see~\cite{MSSM-REV}.}.

In this paper we provide a simple example of a sub-eV Hubble scale
inflation, $H_{\rm inf} \sim 10^{-3}-10^{-1}$~eV, embedded within
MSSM. This is realizable provided supersymmetry (SUSY) breaking is
communicated via gauge mediation, i.e. gauge mediated supersymmetry
breaking (GMSB)~\cite{GMSB}, in contrast to Refs.~\cite{AEGM,AKM}
where we assumed gravity mediation.  We shall predict the reheat
temperature around TeV, which strongly favors electroweak baryogenesis
within MSSM and gravitino as a dark matter candidate. Thus all the
ingredients for a successful cosmology are naturally contained within
MSSM.

Let us first highlight relevant points of the model:

\begin{itemize}

\item{The model is based on inflation in the vicinity of a saddle
point, see~\cite{AEGJM,AM}, which is predictive, radiatively stable,
free from supergravity and Trans-Planckian corrections.  The only
scales which enters in the potential is the weak scale $\sim$ TeV, and
the Grand Unified scale, $M_{\rm GUT}\sim 10^{16.5}$~GeV.}

\item{The reheat temperature of the model is just around TeV, which
points towards thermal electroweak baryogenesis~\cite{BARYO-REV}. If
the reheat temperature is slightly lower than this, we have the right
conditions for a cold electroweak baryogenesis~\cite{GKS}.}

\item{Within GMSB the gravitino is the lightest SUSY particle
(LSP). Therefore a reheat temperature of ${\cal O}({\rm TeV})$ will
lead to a sufficient relic abundance for the (stable) gravitino as a
dark matter candidate~\cite{STEFFEN}.}

\end{itemize}

Let us now consider an MSSM flat direction, ${\phi}$, lifted by a
non-renormalizable ($n > 3$) superpotential term (for a detailed
dynamics on multiple flat direction, see~\cite{EJM}):
\beq\label{supot} 
W = ({\lambda_n}/{n})({\Phi^n}/{M^{n-3}_{\rm
GUT}})\,, 
\eeq
where $\Phi$ is the superfield which contains the flat direction
$\phi$.  Within MSSM all the flat directions are lifted by $n \leq 9$
operators~\cite{DRT,GKM}.  The cut-off scale is $M_{\rm GUT}$,
therefore the above superpotential is a reflection of integrating out
the physics above the GUT scale, and we assume the non-renormalizable
coupling to be $\lambda_{n} \sim {\cal O}(1)$.  Note that the new
physics does not necessarily have to be tied to the GUT physics. For
example a gauged $U(1)_{B-L}$ may appear at an intermediate scale
$M_{B-L} \ll M_{\rm GUT}$. This will amount to the same superpotential
as in Eq.~(\ref{supot}) parameterized by $M_{\rm GUT}$, so long as
$\lambda_n \ll 1$. This is conceivable as $\lambda_n$ is a product of
the Yukawa couplings associated with the new interaction terms beyond
the MSSM.

In GMSB the two-loop correction to the flat direction potential
results in a logarithmic term above the messenger scale, i.e. $\phi >
M_S$~\cite{dGMM}. Together with the $A$-term this leads to the scalar
potential
\beq \label{scpot}
V = M_{F}^4\ln\left(\frac{\phi^2}{M_{S}^2}\right) + 
A\cos(n \theta  + \theta_A)
\frac{\lambda_{n}\phi^n}{n\,M^{n-3}_{\rm GUT}} + \lambda^2_n 
\frac{{\phi}^{2(n-1)}}{M^{2(n-3)}_{\rm GUT}}\,,
\eeq
where $M_F \sim (m_{SUSY} \times M_S)^{1/2}$ and $m_{SUSY} \sim 1$~TeV
is the soft SUSY breaking mass at the weak scale. For $\phi >
M^2_F/m_{3/2}$, usually the gravity mediated contribution, $m^2_{3/2}
\phi^2$, dominates the potential where $m_{3/2}$ is the gravitino
mass.  Here we will concentrate on the VEVs $M_s \ll \phi \ls
M^2_F/m_{3/2}$.

In Eq.~(\ref{scpot}), $\phi$ and $\theta$ denote the radial and the
angular coordinates of the complex scalar field
$\Phi=\phi\,\exp[i\theta]$ respectively, while $\theta_A$ is the phase
of $A$-term (thus $A$ is a positive quantity with a dimension of
mass). Note that the first and third terms in Eq.~(\ref{scpot}) are
positive definite, while the $A$-term leads to a negative contribution
along the directions where $\cos(n \theta + \theta_A) < 0$.  The
cosmological importance of an $A$-term can be found
in~\cite{MSSM-REV,DRT,MARIEKE,CURVATON}.

Although individual terms are unable to support a sub-Planckian VEV
inflation, but as shown in Refs.~\cite{AEGM,AKM,AEGJM,AM}, a
successful inflation can be obtained near the saddle point, which we
find by solving, $V^{\prime}(\phi_0) = V^{\prime\prime}(\phi_0)=0$
(where derivative is w.r.t $\phi$).
\begin{eqnarray}
\label{saddle2}
\phi_0 &=& \left( \frac{M^{n-3}_{\rm GUT} M_F^2}{\lambda_n}
\sqrt{\frac{n}{(n-1)(n-2)}} \right)^{1/(n-1)}\,, \\ 
A &=& \frac{4(n-1)^2
\lambda_n}{n M^{n-3}_{\rm GUT}} \phi_0^{n-2}\,. \label{A2}
\end{eqnarray}
In the vicinity of the saddle point, we obtain the total energy
density and the third derivative of the potential to be:
\begin{eqnarray}
\label{V0}
V(\phi_0) &=& M_F^4 \left[ \ln\left(\frac{\phi_0^2}{M_S^2} \right) -
  \frac{3n-2}{n(n-1)} \right]\,, \\
V^{\prime\prime\prime}(\phi_0)& =& 4n(n-1) M_F^4 \phi_0^{-3}\,.
\end{eqnarray}
There are couple of interesting points, first of all note that the
scale of inflation is extremely low in our case, barring some small
coefficients of order one, the Hubble scale during inflation is given
by:
\beq
\label{Hinf}
H_{\rm inf}\sim {M_{F}^2}/{M_{\rm P}} \sim 10^{-3}-10^{-1}~{\rm eV} \,,
\eeq
for $M_F \sim 1-10$ TeV. For such a low scale inflation usually it is
extremely hard to obtain the right phenomenology. But there are
obvious advantages of having a low scale inflation, $M_F \gg H_{\rm
inf}$. The supergravity corrections and the Trans-Planckian
corrections are all negligible~\cite{AEGJM}, therefore the model
predictions are trustworthy.

Perturbations which are relevant for the COBE normalization are
generated a number ${\cal N}_{\rm COBE}$ e-foldings before the end of
inflation. The value of ${\cal N}_{\rm COBE}$ depends on thermal
history of the universe and the total energy density stored in the
inflaton, which in our case is bounded by, $V_0 \leq 10^{16}~({\rm
GeV})^4$. The required number of e-foldings yields in our case, ${\cal
N}_{\rm COBE}\sim 40$~\cite{BURGESS}, provided the universe
thermalizes within one Hubble time. Although within SUSY
thermalization time scale is typically very long~\cite{AVERDI1},
however, in this particular case it is possible to obtain a rapid
thermalization.

Near the vicinity of the saddle point, $\phi_0$, the potential is
extremely flat and one enters a regime of
self-reproduction~\cite{LINDE}. The self-reproduction regime lasts as
long as the quantum diffusion is stronger than the classical drag;
$H_{\rm inf}/2\pi >\dot\phi/H_{\rm inf}$, for $\phi_s \leq \phi \leq \phi_0$,
where $\phi_0 - \phi_s \simeq M_{F}\left({\phi_0}/{M_{\rm
P}}\right)^{3/2}$.  From then on, the evolution is governed by the
classical slow roll. Inflation ends when $\vert \eta \vert \sim 1$,
which happens at $\phi \simeq \phi_e$, where
\beq
\phi_e-\phi_0=-\sqrt{\frac{\sqrt{2}V_0\phi_0^3}{2n(n-1)M_F^4M_{\rm P}}}\,.
\eeq
Assuming that the classical motion is due to the third derivative of
the potential, $V^{\prime}(\phi)\simeq (1/2)
V^{\prime\prime\prime}(\phi_0)(\phi-\phi_0)^2$, the total number of
e-foldings during the slow roll period is found to be:
\beq
\label{N}
{\cal N}_{tot}=\int_{\phi_s}^{\phi_e}\frac{H_{\rm inf} d\phi}{\dot\phi}\simeq
\frac{2V_0\phi_0^3}{4n(n-1)M_{F}^4M_{\rm P}^2}\Big(\frac{1}{\phi_0-\phi_s}\Big)
\,.
\eeq
%
%
%
This simplifies to 
%
\beq
\label{N1}
{\cal N}_{\rm tot}\simeq {\phi_{0}^{3/2}}/({M_{\rm P}^{1/2}M_F})\,.
\eeq
Let us now consider the adiabatic density perturbations. Despite
$H_{\rm inf}\ll 1$~eV, the flat direction can generate adequate
density perturbations as required to explain to match the
observations. Recall that inflation is driven by $V^{\prime \prime
\prime}\neq 0$, we obtain
\beq
\label{amp}
\delta_{H}\simeq ({1/5\pi})({H^2_{\rm inf}}/{\dot\phi})
\sim {M_F^2 M_{\rm P} {\cal N}_e^2}{\phi_0^{-3}}\sim 10^{-5}\,.
\eeq
Note that for $M_F \sim 10$~TeV, and ${\cal N}_{\rm COBE}\sim 40$, we
match the current observations~\cite{WMAP3}, when $\phi_0\sim
10^{11}$~GeV. The validity of Eq.~(\ref{scpot}) for such a large VEV
requires that $M^2_F > (10^{11}~{\rm GeV}) \times m_{3/2}$. For $M_F
\sim 10$ TeV this yields the bound on the gravitino mass, $m_{3/2} <
1$ MeV, which is compatible with the dark matter constraints as we
will see.

We can naturally satisfy Eq.~(\ref{amp}) provided, $n=6$. The
non-renormalizable operator, $n=6$, points towards two MSSM flat
directions out of many,
\beq
LLe ~~~~~{\rm and}~~~~~ udd\,.
\eeq
As we discussed before in \cite{AEGM}, these are the only directions
which are suitable for inflation as they give rise to a non-vanishing
$A$-term. Note that the inflatons are now the gauge invariant objects.
The total number of e-foldings, during the slow roll inflation, after
using Eq.~(\ref{N1}) yields,
\beq
\label{Ntot}
{\cal N}_{\rm tot}\sim 10^{3}\,.
\eeq
While the spectral tilt and the running of the power spectrum are
determined by ${\cal N}_{\rm COBE} \sim 40 \ll {\cal N}_{\rm tot}$.
\begin{eqnarray}
\label{spect}
& & n_s = 1 + 2\eta - 6\epsilon \simeq 1 - \frac{4}{{\cal N}_{\rm COBE}} 
\sim 0.90\,,\\
\label{runspect}
& & \frac{d\,n_s}{d\ln k} = 16\epsilon\eta - 24 
\epsilon^2 - 2\xi^2 \simeq - \frac{4}{{\cal N}_{\rm COBE}^2} 
\sim - 10^{-3} \,, \nonumber \\
& & \,
\end{eqnarray}
where $\xi^2 = M_P^4 V^{\prime} V^{\prime\prime\prime}/ V^2$. Note
that the spectral tilt is slightly away from the $2\sigma$ result of
the current WMAP 3 years data, on the other hand running of the
spectrum is well inside the current bounds~\cite{WMAP3}.

At first instance one would discard the model just from the slight
mismatch in the spectral tilt from the current observations. However
note that our analysis strictly assumes that the slow roll inflation
is driven by $V^{\prime\prime\prime}(\phi_0)$. This is particularly
correct if $V^{\prime}(\phi_0) =0$ and
$V^{\prime\prime}(\phi_0)=0$. Let us then study the case when
$V^{\prime}(\phi_0)\neq 0$, as discussed in~\cite{AM}.

The latter case can be studied by parameterizing a small deviation
from the exact saddle point condition by solving near the point of
inflection, where we wish to solve $V^{\prime\prime}(\phi_0)=0$ and we
get up-to 1st order in the deviation, $\delta<1$,
\beq
\label{inflection1}
\tilde{A} = A(1-\delta), \qquad \tilde{\phi}_0 = \phi_0 \left( 1 -
  \frac{n-1}{n(n-2)} \delta \right)\,,
\eeq
with $A$ and $\phi_0$ are the saddle point solutions. Then the 1st
derivative is given by
\beq
\label{a1}
V^{\prime}(\phi_0) = 4 \frac{n-1}{n-2} M_F^4 \phi_0^{-1} \delta\,.
\eeq
Therefore the slope of the potential is determined by,
$V^{\prime}(\phi) \simeq V^{\prime}(\phi_0) + (1/2) V^{\prime \prime
\prime}(\phi_0) (\phi - \phi_0)^2$.

Note that both the terms on the right-hand side are positive.  The
fact that $V^{\prime}(\phi_0) \neq 0$ can lead to an interesting
changes from the saddle point behavior, for instance the total number
of e-foldings is now given by
\beq 
\label{tot2}
{\cal N}_{\rm tot} = \frac{V(\phi_0)}{M^2_{\rm P}}
\int_{\phi_{\rm end}}^{\phi_0} \frac{d\phi}{{V^{\prime}(\phi_0) + 
\frac{1}{2} V^{\prime \prime \prime}(\phi_0) (\phi - \phi_0)^2}}\,.
\eeq
First of all note that by including $V^{\prime}$, we are slightly away
from the saddle point and rather close to the point of inflection.
This affects the total number of e-foldings during the slow roll. It
is now much less than that of ${\cal N}_{\rm tot}$, i.e. ${\cal
N}_{\rm tot}\ll 10^3$, see Eq.~(\ref{Ntot}).

When both the terms in the denominator of the integrand contributes
equally then there exists an interesting window.
\beq\label{constr007}
\frac{\kappa}{8}\leq \delta \leq \frac{\kappa}{2}\,.
\eeq
where
\beq\label{kappa} 
\kappa \equiv \frac{n-2}{n(n-1)^2} \left[ \ln\left(
\frac{\phi_0^2}{M_S^2} \right) - \frac{3n-2}{n(n-1)} \right]^2 
\frac{\phi_0^4}{M_{\rm P}^4 {\cal N}_{\rm COBE}^2}\,.
\eeq
The lower limit in Eq.~(\ref{constr007}) is saturated when
$V^{\prime}(\phi_0)=0$, while the upper limit is saturated when ${\cal
N}_{\rm tot}\simeq {\cal N}_{\rm COBE} \simeq 40$. It is also easy to
check that there will be no self-reproduction regime for the field
values determined by $\delta$.

It is a straightforward but a tedious exercise to demonstrate that
when the upper limit of Eq.~(\ref{constr007}) is saturated the
spectral tilt becomes $n_s\simeq 1$, when the lower limit is satisfied
we recover the previous result with $n_s=0.90$. This value, $n_s
\rightarrow 1$, can be easily understood as $\phi_{\rm COBE}
\rightarrow \phi_0$ (where $\phi_{\rm COBE}$ corresponds to the VeV
where the end of inflation corresponds to ${\cal N}_{\rm COBE}\sim
40$), in which case, $\eta \rightarrow 0$.  Therefore the spectral
tilt becomes nearly scale invariant. We therefore find a
range~\cite{AM},
\beq
0.90 \leq n_s\leq 1\,,
\eeq
whose width is within the $2\sigma$ error of the central
limit~\cite{WMAP3}.  Similarly the running of the spectral tilt gets
modified too but remains within the observable limit~\footnote{A
similar exercise can be done for the running of the spectral tilt and
the running lies between $-16/{\cal N}_{\rm COBE}^2 \leq d n_s/d\ln
k\leq -4/{\cal N}_{\rm COBE}^2$~\cite{AM}.}, while the amplitude of
the power spectrum is least affected~\cite{AM}.

Let us now discuss the issue of reheating and thermalization.
Important point is to realize that the inflaton belongs to the MSSM,
i.e. $LLe$ and $udd$, both carry MSSM charges and both have gauge
couplings to gauge bosons and gauginos. After inflation the condensate
starts oscillating.  The effective frequency of the inflaton
oscillations in the Logarithmic potential, Eq.~(\ref{scpot}), is of
the order of $M^2_{F}/\phi_0$, while the expansion rate is given by
$H_{\rm inf} \sim M^2_F/M_{\rm P}$. This means that within one Hubble
time the inflaton oscillates nearly $M_{\rm P}/\phi_0 \sim 10^{7}$
times. The motion of the inflaton is {\it strictly } one dimensional
from the very beginning. During inflation, the imaginary direction is
very heavy and settles down in the minimum of the potential.

An efficient bout of particle creation occurs when the inflaton
crosses the origin, which happens twice in every oscillation. The
reason is that the fields which are coupled to the inflaton are
massless near the point of enhanced symmetry. Mainly electroweak gauge
fields and gauginos are then created as they have the largest coupling
to the flat direction.  The production takes place in a short
interval. Once the inflaton has passed by the origin, the gauge
bosons/gauginos become heavy by virtue of VeV dependent masses and
they eventually decay into particles sparticles, which creates the
relativistic thermal bath. This is so-called instant preheating
mechanism~\cite{INSTANT}.  In a favorable condition, the flat
direction VeV coupled very weakly to the flat direction inflaton could
also enhance the perturbative decay rate of the inflaton~\cite{ABM}.
In any case there is no non-thermal gravitino production~\cite{MAROTO}
as the energy density stored in the inflaton oscillations is too low.

A full thermal equilibrium is reached when ${\it a)~kinetic }$ and
${\it b)~chemical~equilibrium }$ are established~\cite{AVERDI1}. The
maximum temperature of the plasma is given by
\beq
\label{tmax}
T_{\rm R} \sim [V(\phi_0)]^{1/4} \sim M_F \ls 10 {\rm
TeV}\,, 
\eeq
when the flat direction, either $LLe$ or $udd$ evaporates completely.
This naturally happens at the weak scale. There are two very important
consequences which we summarize below.

{\it Hot or cold electroweak Baryogenesis}:~The model strongly favors
electroweak baryogenesis within MSSM.  Note that the reheat
temperature is sufficient enough for a thermal electroweak
baryogenesis~\cite{BARYO-REV}.

However, if the thermal electroweak baryogenesis is not triggered,
then cold electroweak baryogenesis is still an option~\cite{GKS}.
During the cold electroweak baryogenesis, the large gauge field
fluctuations give rise to a non-thermal sphaleron transition.  In our
case it is possible to excite the gauge fields of $SU(2)_{L}\times
U(1)_{Y}$ during instant preheating provided the inflaton is $LLe$.
The $LLe$ as an inflaton carries the same quantum number which has a
$B-L$ anomaly and large gauge field excitations can lead to
non-thermal sphaleron transition to facilitate baryogenesis within
MSSM.

{\it Gravitino dark matter}:~Within GMSB gravitinos are the LSP and if
the $R$-parity is conserved then they are an excellent candidate for
the dark matter. There are various sources of gravitino production in
the early universe~\cite{STEFFEN,AHJMP}. However in our case the
thermal production is the dominant one and mainly helicity $\pm 1/2$
gravitinos are created. Gravitinos thus produced have the correct
dark matter abundance for~\cite{dGMM,BUCH}
\beq \label{dm}
{m_{3/2} \over 100~ {\rm keV}} \simeq {1 \over {\rm few}} \Big({T_{\rm R} 
\over 1~{\rm TeV}}\Big) \Big({M_{\tilde g} \over 1~{\rm TeV}}\Big)^2,
\eeq
where $M_{\tilde g}$ is the gluino mass. For $m_{3/2} \gs 100$ keV, 
Eq.~(\ref{dm}) is easily satisfied for $M_{\tilde g} \sim 1$ TeV and
$T_{\rm R} \ls 10$ TeV. We remind that for ${\cal
O}({\rm keV}) \ls m_{3/2} < 100$ keV gravitinos produced from the
sfermion decays overclose the universe~\cite{dGMM}.

Before concluding we should also highlight that the existence of a
saddle point does not get spoiled through radiative corrections, see
Ref.~\cite{AEGJM}. To summarize, we provided a truly low scale
inflation model with $H_{\rm inf}\sim 10^{-3}-10^{-1}$~eV, embedded
within MSSM, provided GMSB is the correct paradigm. Although inflation
occurs at such low scales, the model predictions match the current
WMAP data and the reheat temperature of $T_{\rm R}\ls 10$ TeV is
sufficient enough to trigger either hot or cold electroweak
baryogenesis. The model also produces sufficient abundance of
gravitinos to be the dark matter candidate.  Thus inflation within
GMSB connects the physics of microwave background radiation to a
successful dark matter and a baryogenesis scenario whose ingredients
are testable at the LHC.

We wish to thank Alex Kusenko and Misha Shaposhnikov for helpful
discussion. The research of RA was supported by Perimeter Institute
for Theoretical Physics. Research at Perimeter Institute is supported
in part by the Government of Canada through NSERC and by the provine
of Ontario through MEDT. The research of AM is partly supported by the
European Union through Marie Curie Research and Training Network
``UNIVERSENET'' (MRTN-CT-2006-035863).



\begin{thebibliography}{99}

\bibitem{AEGM}
R.~Allahverdi, K.~Enqvist, J.~Garcia-Bellido and A.~Mazumdar,
arXiv:hep-ph/0605035.

\bibitem{AKM}
R.~Allahverdi, A.~Kusenko and A.~Mazumdar,
  arXiv:hep-ph/0608138.

\bibitem{AM}
R. Allahverdi and A. Mazumdar, arXiv:hep-ph/0610069,

\bibitem{AEGJM}
R.~Allahverdi, K.~Enqvist, J.~Garcia-Bellido, A.~Jokinen and A.~Mazumdar,
  arXiv:hep-ph/0610134.


\bibitem{FEW}
G.~Lazarides and Q.~Shafi,
  Phys.\ Lett.\ B {\bf 308}, 17 (1993).
S.~Kasuya, T.~Moroi and F.~Takahashi,
  Phys.\ Lett.\ B {\bf 593}, 33 (2004).
 R.~Brandenberger, P.~M.~Ho and H.~C.~Kao,
  JCAP {\bf 0411}, 011 (2004).
A.~Jokinen and A.~Mazumdar,
  Phys.\ Lett.\ B {\bf 597}, 222 (2004).

\bibitem{LYTH1}
 J.~C.~B.~Sanchez, K.~Dimopoulos and D.~H.~Lyth,
  arXiv:hep-ph/0608299.



\bibitem{MSSM-REV}
For reviews, see
 K.~Enqvist and A.~Mazumdar,
  Phys.\ Rept.\  {\bf 380}, 99 (2003);
 M. Dine and A. Kusenko, Rev. Mod. Phys. {\bf 76}, 1 (2004).

\bibitem{GMSB}
G.~F.~Giudice and R.~Rattazzi,
  Phys.\ Rept.\  {\bf 322}, 419 (1999).

\bibitem{BARYO-REV}
V.~A.~Rubakov and M.~E.~Shaposhnikov,
  Usp.\ Fiz.\ Nauk {\bf 166}, 493 (1996)
  [Phys.\ Usp.\  {\bf 39}, 461 (1996)];


\bibitem{GKS}
J.~Garcia-Bellido, D.~Y.~Grigoriev, A.~Kusenko and M.~E.~Shaposhnikov,
  Phys.\ Rev.\ D {\bf 60} (1999) 123504.


\bibitem{STEFFEN}
F.~D.~Steffen,
  arXiv:hep-ph/0605306.

\bibitem{EJM}
 K.~Enqvist, A.~Jokinen and A.~Mazumdar,
  JCAP {\bf 0401}, 008 (2004).


\bibitem{DRT}
M.~Dine, L.~Randall and S.~Thomas, Phys. Rev. Lett. {\bf 75}, 398 (1995).
M. Dine, L. Randall and S. Thomas, Nucl. Phys. B {\bf 458}, 291 (1996).

\bibitem{GKM}
T. Gherghetta, C. Kolda and S. P. Martin, Nucl. Phys. B {\bf 468},
37 (1996).

\bibitem{dGMM}
  A.~de Gouvea, T.~Moroi and H.~Murayama,
  Phys.\ Rev.\ D {\bf 56}, 1281 (1997).



\bibitem{MARIEKE}
M. Postma and A. Mazumdar, 
JCAP {\bf 0401}, 005 (2004).


\bibitem{CURVATON}
 R.~Allahverdi, K.~Enqvist, A.~Jokinen and A.~Mazumdar,
  arXiv:hep-ph/0603255.



\bibitem{BURGESS}
C.~P.~Burgess, R.~Easther, A.~Mazumdar, D.~F.~Mota and T.~Multamaki,
  JHEP {\bf 0505}, 067 (2005).

\bibitem{AVERDI1}
 R.~Allahverdi and A.~Mazumdar,
  arXiv:hep-ph/0505050.
R.~Allahverdi and A.~Mazumdar,
  arXiv:hep-ph/0512227.
 R.~Allahverdi and A.~Mazumdar,
  arXiv:hep-ph/0603244.
 R.~Allahverdi and A.~Mazumdar,
  arXiv:hep-ph/0608296.

         

\bibitem{LINDE}
A. D. Linde, PARTICLE PHYSICS AND INFLATIONARY COSMOLOGY (Harwood Academic Publishers, Chur, Switzerland 1990).


\bibitem{WMAP3}
D.N. Spergel, et.al., astro-ph/0603449.


\bibitem{INSTANT}
G.~N.~Felder, L.~Kofman and A.~D.~Linde,
  Phys.\ Rev.\ D {\bf 59}, 123523 (1999).




\bibitem{ABM}
R.~Allahverdi, R.~Brandenberger and A.~Mazumdar,
  Phys.\ Rev.\ D {\bf 70}, 083535 (2004)
  [arXiv:hep-ph/0407230].




\bibitem{MAROTO}
 A.~L.~Maroto and A.~Mazumdar,
  Phys.\ Rev.\ Lett.\  {\bf 84}, 1655 (2000).
 



\bibitem{AHJMP}
R.~Allahverdi, S.~Hannestad, A.~Jokinen, A.~Mazumdar and S.~Pascoli,
  arXiv:hep-ph/0504102.


\bibitem{BUCH}
M.~Bolz, A.~Brandenburg and W.~Buchm\"uller,
Nucl. Phys. B {\bf 606}, 518 (2001).
  

  



\end{thebibliography}
\end{document}